\documentclass[aps, pra, floatfix, twocolumn, superscriptaddress, showpacs, 10pt]{revtex4-1}
\usepackage{amssymb, amsmath, color, mciteplus, graphicx, subfigure}
\usepackage{calc, epsfig, epstopdf, color, mciteplus, bm, mathrsfs}
\usepackage{times}
\usepackage{float}
\usepackage{lipsum}

\begin{document}

\title{Accelerated Super-robust Nonadiabatic Holonomic Quantum Gates}

\author{Pu Shen}

\author{Yan Liang}
\affiliation{Key Laboratory of Atomic and Subatomic Structure and Quantum Control (Ministry of Education), and School of Physics, South China Normal University, Guangzhou 510006, China}

\author{Tao Chen}

\author{Zheng-Yuan Xue}\email{zyxue83@163.com}
\affiliation{Key Laboratory of Atomic and Subatomic Structure and Quantum Control (Ministry of Education), and School of Physics, South China Normal University, Guangzhou 510006, China}
\affiliation{Guangdong Provincial Key Laboratory of Quantum Engineering and Quantum Materials, Guangdong-Hong Kong Joint Laboratory of Quantum Matter,  and Frontier Research Institute for Physics,\\ South China Normal University, Guangzhou 510006, China}

\date{\today}

\begin{abstract}
The nonadiabatic holonomic quantum computation based on three-level systems has wide applicability experimentally due to its simpler energy level structure requirement and inherent robustness from the geometric phase. However, in previous conventional schemes, the states of the calculation subspace have always leaked to the non-computation subspace, resulting in less robustness than anticipated. Recent efforts to address this problem are at the cost of excessively long gate time, which will lead to more decoherence-induced errors. Here, we propose a solution to the problem without the severe limitation of the much longer gate-tine. Specifically, we implement arbitrary holonomic gates via a three-segment Hamiltonian, where the gate time depends on the rotation angle, and the smaller the rotation angle, the shorter the gate time will be. Compared with the previous solutions, our numerical simulations indicate that the decoherence-induced gate errors of our scheme are greatly decreased and the robustness of our scheme is also better, particularly for small-angle rotation gates. Moreover, we provide a detailed physical realization of our proposal on a two-dimensional superconducting quantum circuit. Therefore, our protocol provides a promising alternative for future fault-tolerant quantum computation.
\end{abstract}

\maketitle

\section{Introduction}
Quantum computers have the potential to solve some of the hard problems for classical computers \cite{1}. Meanwhile, the superiority of quantum computers has been proved experimentally \cite{frank2019,zhong2020,wu2021}, and thus the research of quantum computation has attracted much attention in the last several decades. To build a universal quantum computer, a set of high-fidelity and robust universal gates is necessary \cite{david1985}. To this end, a large amount of research has been carried out in a variety of physical systems \cite{2,3,4,5}, among which the development of superconducting circuit systems is quite rapid in the current \cite{6,7,8,9,10}. However, the physical implementation of large-scale  quantum computers is still affected by the decoherence effect and operational-induced errors. Therefore, utilizing geometric phases \cite{berry1984,simon1983,wilczek1984,aharonov1987,anandan1988} to construct gates has attracted researchers' attention \cite{zanardi1999,wang2001,zhu2002,zhu2003,sjoqvist2008,xu2012,sjoqvist2012}, because it only depends on the overall evolution path but not on the specific evolution details, and thus provides an  alternative for high-fidelity and robust universal gates. In the beginning, the Abelian geometric phase was used to implement quantum computation \cite{zhu2002,zhu2003}, which was subsequently extended to exploit the non-Abelian geometric phase due to its non-commutative characteristic\cite{xu2012,sjoqvist2012}. And the nonadiabatic holonomic quantum computation (NHQC) \cite{xu2012,sjoqvist2012} based on the non-Abelian geometric phase is typical because of the shorter gate time than the adiabatic case. Moreover, the NHQC's robust performance has been verified theoretically \cite{chen2018,liu2019, lisai2020, xujing2020, zhoujian2021, lisai2021} and experimentally \cite{yan2019, zhu2019, xu2018, ai2020, ai2022}.

However, in the traditional NHQC schemes \cite{xu2012,sjoqvist2012}, the states of the calculation subspace have always leaked to the non-computation subspace, resulting in less robustness than expected. To overcome  this drawback, it  was proposed \cite{liu2021} to impose a super-robust condition, but this solution can only be used to realize the $\pi$-angle rotation gates. Later, in another experimental work \cite{li2021}, to realize arbitrary rotation gates, a super-robust nonadiabatic holonomic quantum computation (SR-NHQC) scheme was realized by six-segment Hamiltonian, where the quantum gates have fourth-order resistance to the global control error, which is greatly improved compared with the conventional single-loop NHQC schemes \cite{sjiqvist2016,hong2018}. However, this scheme has a shortcoming, that is, the gate time is twice that of the single-loop  scheme \cite{sjiqvist2016,hong2018}, and is independent of the rotation angle. This shortcoming leads to poor gate performance when taking the decoherence effect into consideration, which is quite fatal in practical applications. Therefore, this paper introduces a method to accelerate the SR-NHQC scheme, termed the ASR-NHQC, to solve the problem of excessively long gate time.

Our scheme is based on a detuned $\Lambda$-type three-level system by piecewise Hamiltonian to realize arbitrary holonomic quantum gates. Compared with the previous SR-NHQC scheme \cite{li2021}, our scheme has the following advantages. Firstly, it only needs a three-segment Hamiltonian to construct arbitrary rotation gates, which is a six-segment Hamiltonian in the previous scheme \cite{li2021}. Secondly, our gate time depends on the rotation angle. The smaller the rotation angle, the shorter the gate time, even our longest gate time is much shorter than the previous scheme. Through numerical simulations, we found that our scheme has a great improvement in decoherence performance compared to previous schemes because of the shortening of our gate time. Finally, when considering the coupling strength and the detuning errors, all errors can be classified into these two types of error,   our scheme is more robust than the previous scheme, especially in small-angle rotation gates.

Furthermore, we propose a comprehensive physical realization of a two-dimensional (2D) superconducting quantum circuit for the ASR-NHQC. From our numerical simulations, the single-qubit gate fidelity can be above $99.50\%$ and the two-qubit  $CZ$ gate fidelity is $99.30\%$. Besides,   two-qubit gates here are constructed by only two physical qubits, which is a new implementation on superconducting quantum circuits. Therefore, our scheme provides a fast and implementable alternation for the SR-NHQC and thus is promising  for future  fault-tolerant quantum computation.

\begin{figure}[tbp]
	\includegraphics[width=0.9\linewidth]{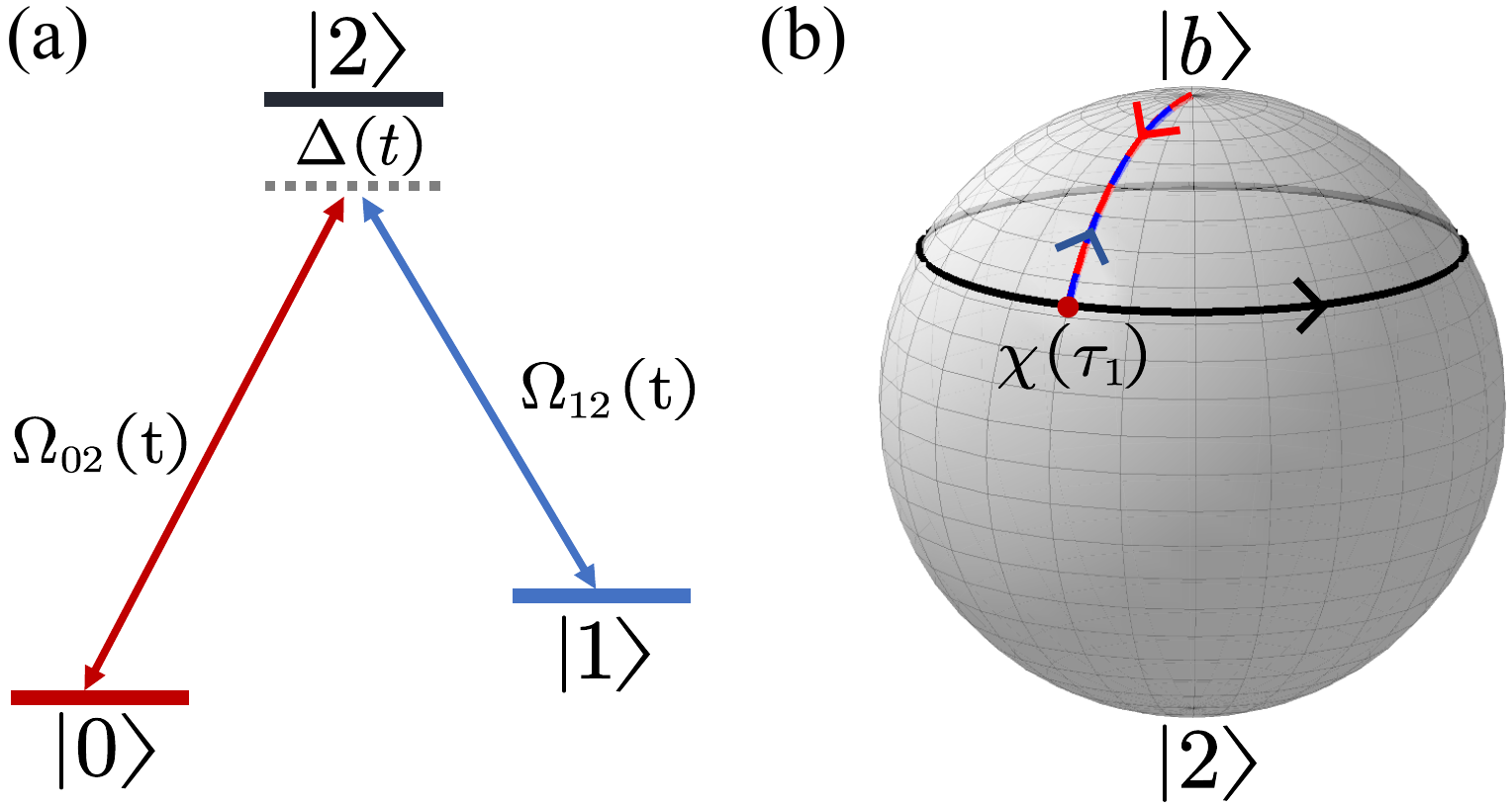}
	\caption{Illustration of our scheme. (a) A $\Lambda$-type three-level system with detuning, where two pulses $\Omega_{02}(t)$ and $\Omega_{12}(t)$ drive $|0\rangle\leftrightarrow|2\rangle$ and $|1\rangle\leftrightarrow|2\rangle$ respectively. (b) Description of the $|\tilde{\psi}_2(t)\rangle$ evolution path on the Bloch sphere composed of $\{|b\rangle,|2\rangle\}$, where $\chi(t)$ and $\zeta(t)$ represent the change of polar angle and azimuth angle respectively. In the beginning, $|\tilde{\psi}_2(0)\rangle$ starts from the North Pole, evolves along the meridian ( red dashed line), and arrives at $\chi(\tau_1)$ at time $\tau_1$, then evolves along the latitude (black solid line) for a circle and returns to $\chi(\tau_1)$ at time $\tau_2$, and finally, evolves along the meridian ( blue dashed line) back to the North Pole at time $\tau$.}
	\label{Fig.1}
\end{figure}

\section{Super-robust holonomic quantum gates}
We consider a Hilbert space of ($M\!+\!N$)-dimensions, spanned by $\{|\psi_k(0)\rangle\}^{M+N}_{k=1}$, where the $M$- and $N$-dimensional subspaces belong to the computational subspace and the non-computation subspace, respectively, and $|\psi_k(t)\rangle$ satisfy the Schr\"{o}dinger equation $i|\dot{\psi}_k(t)\rangle \!=\! \mathcal{H}(t)|\psi_k(t)\rangle$, setting $\hbar \!=\! 1$ hereafter. The time-evolution operator is $U(t) \!=\! \mathcal{T}e^{-i\int^t_0 \mathcal{H}(t')dt'}$, with  $\mathcal{T}$ being the time-ordering operator. The traditional NHQC scheme \cite{xu2012,sjoqvist2012} works as follows. Choosing a different set of time-dependent auxiliary basis vectors $\{|\tilde{\psi_k}(t)\rangle\}^{M}_{k=1}$ in the computation subspace,  the $|\psi_k(t)\rangle$ in the computation subspace can be expressed by $\{|\tilde{\psi}_m(0)\rangle\}^{M}_{m=1}$, i.e., $|\psi_k(t)\rangle \!=\! \sum^M_{m=1}C_{mk}(t)|\tilde{\psi}_m(t)$, with $C_{mk}(t)$ being time dependent coefficients. When the $|\psi_k(t)\rangle$ in the computation subspace meets the cyclic evolution condition and the parallel transport condition, i.e.,
\begin{gather}
\sum_{k=1}^M{|\psi_k(\tau)\rangle\langle\psi_k(\tau)|} = \sum_{k=1}^M{|\psi_k(0)\rangle\langle\psi_k(0)|}, \\
    \langle\psi_k(t)|\mathcal{H}(t)|\psi_l(t)\rangle = 0,\quad k,l = 1,...,M,
\end{gather}
with $\tau$ being the evolution time,  the obtained operator  
\begin{equation}
    U(\tau) = \sum^M_{m,l=1} [\mathcal{T}e^{i\int^\tau_0{\text{A}(t)dt}}]_{ml}|\tilde{\psi}_m(0)\rangle\langle\tilde{\psi}_l(0)|,
    \label{u0}
\end{equation}
is a holonomic matrix acting on the computation subspace, where $\text{A}_{ml}(t) \!=\! i\langle\tilde{\psi}_m(t)|d/dt|\tilde{\psi}_l(t)\rangle$ and $|\tilde{\psi}_m(\tau)\rangle \!=\! |\tilde{\psi}_m(0)\rangle \!=\! |\psi_m(0)\rangle$.

Now, we introduce a global static control error to the system, i.e., $\mathcal{H}_{er}(t) = (1+\iota)\mathcal{H}(t)$ with $\iota$ being a small error fraction. The evolution operator with error can be expressed as $U_{er}(\tau) \!=\! \mathcal{T}e^{-i\int^\tau_0 \mathcal{H}_{er}(t)dt}$, which is difficult to solve analytically. But, with the Magnus expansion \cite{ribeiro2017,magnus1954,blanes2009}, we can perturbatively process the evolution operator $U_{er}(\tau) \!=\! \exp{\sum^\infty_{k=1}{\Lambda_k(\tau)}}$, where $\Lambda_k$ denotes the terms of the Magnus expansion \cite{liu2021}. In this way, we can obtain  the gate fidelity \cite{souza2012,genov2017} under the error as
\begin{align}
    F &= \frac{1}{M}|\text{Tr}[U(\tau)U^\dag_{er}(\tau)]| \notag \\
      &\approx 1-\frac{\iota^2}{2M}\sum^M_{m=1}\sum^{M+N}_{k=1}|D_{mk}|^2-\mathcal{O}(\iota^4),
    \label{F0}
\end{align}
where $D_{mk} \!=\! \int_0^{\tau}{\langle\psi_m(t)|\mathcal{H}(t)|\psi_k(t)\rangle}dt$ with $m \!\in\! M$ and $k \!\in\! M\!+\!N$. For the SR-NHQC schemes \cite{liu2021,li2021}, they demand $D_{mk} \!=\! 0$, and thus the quantum gates have fourth-order resistance to this  error. However, we note that, since the parallel transport condition leads to $D_{mk} \!=\! 0$ with $m,k \!\in\! M$,   the  super-robust condition can be simplified as
\begin{equation}
    \int_0^{\tau}{\langle\psi_k(t)|\mathcal{H}(t)|\psi_l(t)\rangle}dt = 0,\quad k\in M,l\in N.
    \label{sr}
\end{equation}
Moreover, the super-robust condition can suppress the coupling of states in the computational and non-computational subspaces, thus further improving the implemented gate robustness, which has been verified  experimentally \cite{li2021}. However, the previous SR-NHQC scheme \cite{li2021} requires twice the gate time than the single-loop NHQC scheme \cite{sjiqvist2016,hong2018}, which leads to poor decoherence performance, so it is not conducive to the application of future experiments.

Next, we propose a scheme to overcome the long gate time drawback, with a  three-level system, consisting of $\{|0\rangle,|1\rangle,|2\rangle\}$, considered. In this case, three orthogonal auxiliary basis vectors can be set as
\begin{align}
    |\tilde{\psi}_1(t)\rangle &= \cos\frac{\theta}{2}|0\rangle+\sin\frac{\theta}{2}e^{i\varphi}|1\rangle,\notag \\
    |\tilde{\psi}_2(t)\rangle &= \cos\frac{\chi(t)}{2}\sin\frac{\theta}{2}e^{-i\varphi}|0\rangle \notag \\
      & \quad-\cos\frac{\chi(t)}{2}\cos\frac{\theta}{2}|1\rangle+\sin\frac{\chi(t)}{2}e^{i\zeta(t)}|2\rangle,\notag \\
    |\tilde{\psi}_3(t)\rangle &= \sin\frac{\chi(t)}{2}\sin\frac{\theta}{2}e^{-i[\varphi+\zeta(t)]}|0\rangle \notag \\
      & \quad-\sin\frac{\chi(t)}{2}\cos\frac{\theta}{2}e^{-i\zeta(t)}|1\rangle-\cos\frac{\chi(t)}{2}|2\rangle,
\end{align}
where $\theta$ and $\varphi$ are constants, and $\chi(t)$ and $\zeta(t)$ are time-dependent parameters with $\chi(\tau) \!=\! \chi(0) \!=\! 0$ and $\zeta(0) \!=\! \zeta_0$, where $\zeta_0$ can take any value. Besides, we set  $|\tilde{\psi}_1(t)\rangle \!=\! |\psi_1(t)\rangle$ is decoupled from the system and $|\tilde{\psi}_k(t)\rangle \!=\! e^{-i\gamma_k(t)}|\psi_k(t)\rangle$ with $k=2,3$. Here,  the target Hamiltonian is reversely constructed by auxiliary basis vectors \cite{zhao2020}, i.e.,
\begin{align}
    \mathcal{H}(t) &= \left[i\langle \tilde{\psi}_2(t)|d/dt| \tilde{\psi}_{3}(t)\rangle |\tilde{\psi}_2(t)\rangle \langle\tilde{\psi}_{3}(t)|+\text{H.c.}\right] \notag \\
    &+\left[ i\langle \tilde{\psi}_{3}(t)| d/dt |\tilde{\psi}_{3}(t)\rangle -\dot{\gamma}_3(t) \right] |\tilde{\psi}_{3}(t)\rangle \langle \tilde{\psi}_{3}(t)|.
\end{align}
Setting $\dot{\gamma}_3(t) \!=\! \left[ 3\!+\!\cos \chi(t) \right]\dot{\zeta}(t)/2$, we obtain a $\Lambda$-type three-level system with detuning, as shown in Fig. \ref{Fig.1}(a), which is controlled by
\begin{equation}
        \mathcal{H}(t) = \Delta(t)|2\rangle\langle2|+\left[\Omega_{02}(t)|2\rangle\langle0|+\Omega_{12}(t)|2\rangle\langle1|+\text{H.c.}\right],
        \label{H0}
\end{equation}
where $\Omega_{02}(t) \!=\! \Omega(t)\sin(\theta/2)e^{i\varphi}$ and $\Omega_{12}(t) \!=\! -\Omega(t)\cos(\theta/2)$ drive $|0\rangle \! \leftrightarrow \! |2\rangle$ and $|1\rangle \! \leftrightarrow \! |2\rangle$ respectively with the detuning $\Delta(t) \!=\! -\dot{\zeta}(t)\left[ 1\!+\!\cos \chi(t) \right]$, in which $\Omega(t) \!=\! \left[i\dot{\chi}(t) \!+\!\dot{\zeta}(t) \sin \chi(t)\right]\exp[i\zeta(t)]/2$. 

By  defining a bright state $|b\rangle \!=\! \sin(\theta/2)e^{-i\varphi}|0\rangle\!-\!\cos(\theta/2)|1\rangle$, the Hamiltonian can  be rewritten  as
\begin{equation}
    \mathcal{H}(t) = \Delta(t)|2\rangle\langle2|+\left[\Omega(t)|2\rangle\langle b|+\text{H.c.}\right],
\end{equation}
where there exists a dark state $ |d\rangle \!=\! |\tilde{\psi}_1(t)\rangle \!=\! \cos(\theta/2)|0\rangle\!+\!\sin(\theta/2)e^{i\varphi}|1\rangle$ decoupled from the system. Due to the inverse construction, after the evolution period, the time evolution operator acting on the computation subspace is
\begin{align}
    U(\tau) &= |d\rangle\langle d| + e^{-i\gamma}|b\rangle\langle b|\notag \\
    &= \exp{(i\frac{\gamma}{2} \mathbf{n} \cdot \mathbf{\sigma})},
\end{align}
where $\gamma  \!=\! -\gamma_2(\tau) \!=\! \frac{1}{2}\int_0^{\tau}{\left[ 1\!-\!\cos \chi (t) \right] \dot{\zeta}(t) dt}$ is the rotation angle, $\mathbf{n} \!=\! (\sin\theta \cos\varphi,\sin\theta \sin\varphi,\cos\theta)$ is a unit vector, and $\mathbf{\sigma} \!=\! (\sigma_x,\sigma_y,\sigma_z)$ are Pauli operators of $\{|0\rangle,|1\rangle\}$. Obviously, $U(\tau)$ is an arbitrary nonadiabatic holonomic quantum gate by choosing specific parameters $\{\theta,\varphi,\gamma\}$.

For examples, the $H$, $S$, and $T$ gates can be constructed by setting $\{\theta,\varphi,\gamma\} \!=\! \{\pi/4,0,\pi\}, \{0,0,\pi/2\}, \{0,0,\pi/4\}$ respectively. Subsequently, applying the super-robust condition in Eq. (\ref{sr}), we demand $\int_0^{\tau}\!{\langle\psi_k(t)|\mathcal{H}(t)|\psi_3(t)\rangle}dt \!=\! 0$ ($k \!=\! 1,2$). Since $|\psi_1(t)\rangle$ is decoupled from the system, the super-robust condition is reduced to
\begin{align}
    &\int_0^{\tau}{\langle\psi_2(t)|\mathcal{H}(t)|\psi_3(t)\rangle} \notag \\
    &=\int_0^{\tau}{\frac{1}{2}e^{-i\left[ \gamma_2(t) -\gamma_3(t) \right]}e^{-i\zeta(t)}\left[ i\dot{\chi}(t) +\dot{\zeta}(t) \sin \chi(t) \right] dt} \notag \\
    &=\int_0^{\tau}{\frac{1}{2}e^{i\zeta(t)}\left[ i\dot{\chi}(t) +\dot{\zeta}(t) \sin \chi(t) \right] dt} =0,
    \label{srs}
\end{align}
which is a specific parameter requirement for our scheme to implement SR-NHQC. In order to satisfy Eq. (\ref{srs}), we divide the Hamiltonian into the following three segments
\begin{align}
    \mathcal{H}^1(t) &=  \frac{1}{2}i\dot{\chi}(t)e^{i\zeta_0}|2\rangle\langle b|+\text{H.c.}, \quad t\in[0,\tau_1], \notag \\
    \mathcal{H}^2(t) &=  -\dot{\zeta}(t)[1+\cos{\chi(\tau_1)}]|2\rangle\langle 2| \notag \\
    &+\left[\frac{1}{2}\dot{\zeta}(t) \sin{\chi(\tau_1)} e^{i\zeta(t)}|2\rangle\langle b|+\text{H.c.}\right], \quad t\in(\tau_1,\tau_2], \notag \\
    \mathcal{H}^3(t) &=  \frac{1}{2}i\dot{\chi}(t)e^{i\zeta(\tau_2)}|2\rangle\langle b|+\text{H.c.}, \quad t\in(\tau_2,\tau],
    \label{3H}
\end{align}
through the following parameter settings
\begin{align}
    \dot{\zeta}(t) &= 0, \,\zeta(t) = \zeta_0, \quad t\in[0,\tau_1], \notag \\
    \dot{\chi}(t) &= 0, \,\chi(t) = \chi(\tau_1), \,\zeta(\tau_2)-\zeta(\tau_1) = 2\pi, \quad t\in(\tau_1,\tau_2], \notag \\
    \dot{\zeta}(t) &= 0, \,\zeta(t) = \zeta(\tau_2) = 2\pi+\zeta_0, \quad t\in(\tau_2,\tau],
\end{align}
where the choice of $\zeta(t)$ in $\mathcal{H}^2(t)$ determines the mathematical expression of the detuning and the coupling strength. It is worth noting that there is a time-independent detuning $\Delta \!=\! 2\pi[1\!+\!\cos{\chi(\tau_1)}]/(\tau_2\!-\!\tau_1)$ if the $\zeta(t)$ in $\mathcal{H}^2(t)$ is chosen as $\zeta(t) \!=\! 2\pi t/(\tau_2\!-\!\tau_1)\!+\!\zeta_0$. Under the control of the piecewise Hamiltonian of Eq. (\ref{3H}), by simple calculation, we obtain the rotation angle $\gamma=\pi[1-\cos{\chi(\tau_1)}]$ depending on $\chi(\tau_1)$, and the evolutionary trajectory of $|\tilde{\psi}_2(t)\rangle$ can be depicted on the Bloch sphere composed of $\{|b\rangle,|2\rangle\}$, as shown in   Fig. \ref{Fig.1}(b), where $\chi(t)$ and $\zeta(t)$ represent the change of polar angle and azimuth angle respectively. In the beginning, $|\tilde{\psi}_2(0)\rangle$ starts from the North Pole, evolves along the meridian (red dotted line), and arrives at $\chi(\tau_1)$ at time $\tau_1$, then evolves along the latitude (black solid line) for a circle and returns to $\chi(\tau_1)$ at time $\tau_2$, and finally, evolves along the meridian (blue dotted line) back to the North Pole at time $\tau$.

\section{Acceleration and Gate performance}
\begin{figure}[tbp]
	\includegraphics[width=\linewidth]{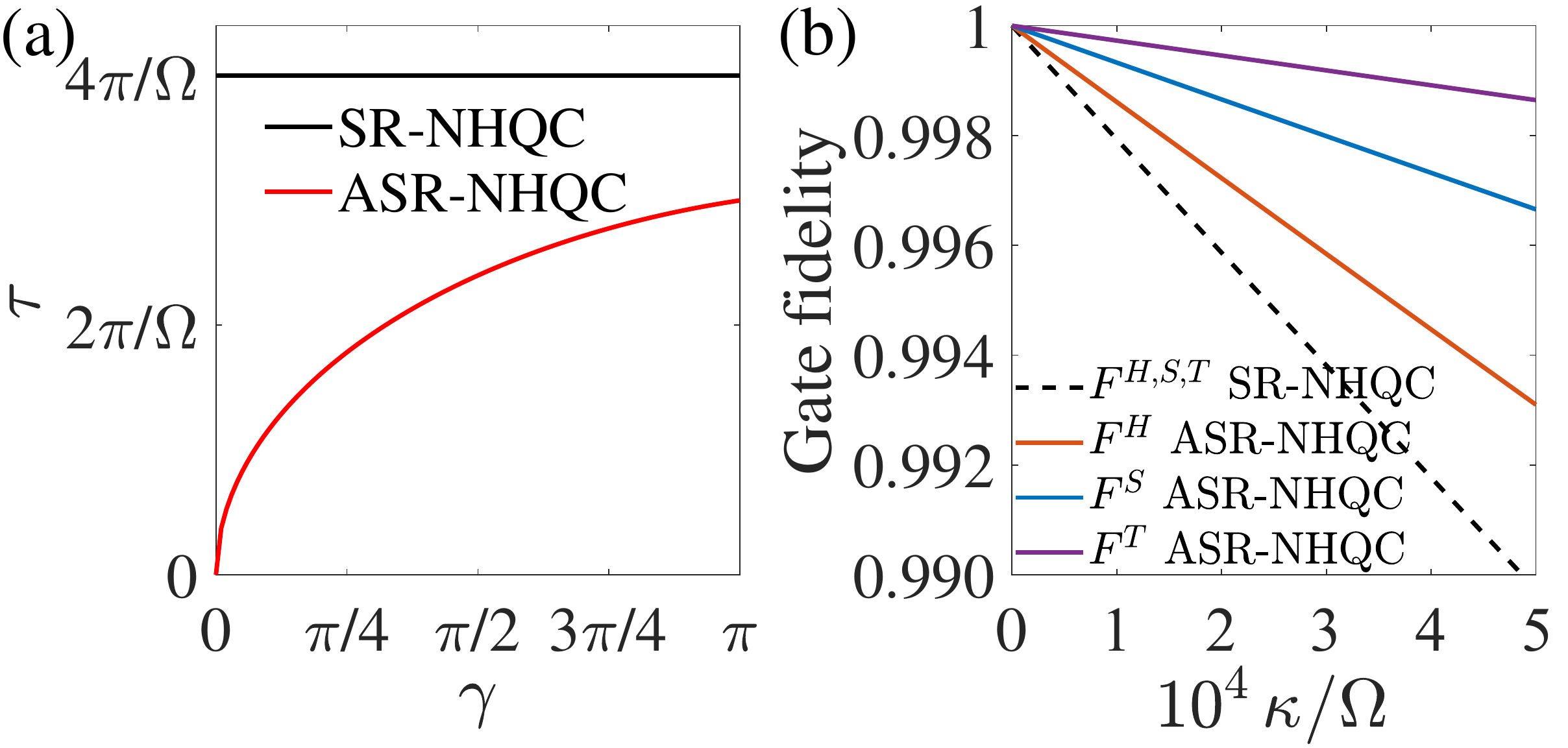}
	\caption{Numerical evaluation of the gate performance. (a) Variation trend of the gate time $\tau$ with the rotation angle $\gamma$ for our scheme (red line) and the previous SR-NHQC (black line). (b) How gate fidelity is affected by the decoherence rate for the $H$, $S$, and $T$ gates. Different gates in our scheme have different effects (three different colors' solid lines) due to different gate times, but different gates in the previous scheme have only the same impact results (dashed line) due to the same gate time.}
	\label{Fig.2}
\end{figure}
In order to quickly find the correlation between the gate time $\tau$ and the rotation angle $\gamma$, we set the coupling strength's magnitude to the same fixed value for each Hamiltonian, i.e.
\begin{align}
    \mathcal{H}^1(t) &=  \frac{1}{2}i\Omega e^{i\zeta_0}|2\rangle\langle b|+\text{H.c.}, \quad t\in[0,\tau_1], \notag \\
    \mathcal{H}^2(t) &=  -\dot{\zeta}(t)[1+\cos{\chi(\tau_1)}]|2\rangle\langle 2| \notag \\
    &+\left[\frac{1}{2}\Omega e^{i\zeta(t)}|2\rangle\langle b|+\text{H.c.}\right], \quad t\in(\tau_1,\tau_2], \notag \\
    \mathcal{H}^3(t) &=  \frac{1}{2}i\Omega e^{i\zeta(\tau_2)}|2\rangle\langle b|+\text{H.c.}, \quad t\in(\tau_2,\tau],
    \label{3Hc}
\end{align}
and thus the operation time of each Hamiltonian can be calculated as $\tau_1 = \chi(\tau_1)/\Omega$, $\tau_2-\tau_1 = 2\pi\sin{\chi(\tau_1)}/\Omega$, $\tau-\tau_2 = \chi(\tau_1)/\Omega$.
Subsequently, the total gate time is expressed as
\begin{equation}
    \tau = \frac{2\arccos{(1-\gamma/\pi)}+2\pi\sin{\left[\arccos{(1-\gamma/\pi)}\right]}}{\Omega},
\end{equation}
where the rotation angle $\gamma\!\in\![0,\pi]$ and thus the gate time $\tau\!\in\![0,3\pi/\Omega]$. However, the gate time is a fixed value $4\pi/\Omega$ for the previous SR-NHQC \cite{li2021} (see Appendix A for details), which is twice the gate time of the single-loop NHQC schemes \cite{sjiqvist2016,hong2018}. In Fig. \ref{Fig.2}(a), we depict numerically the gate time $\tau$ versus the rotation angle $\gamma$ for our scheme and the previous SR-NHQC. Obviously, our scheme greatly shortens the gate time compared to the previous SR-NHQC.

\begin{figure}[tbp]
	\includegraphics[width=\linewidth]{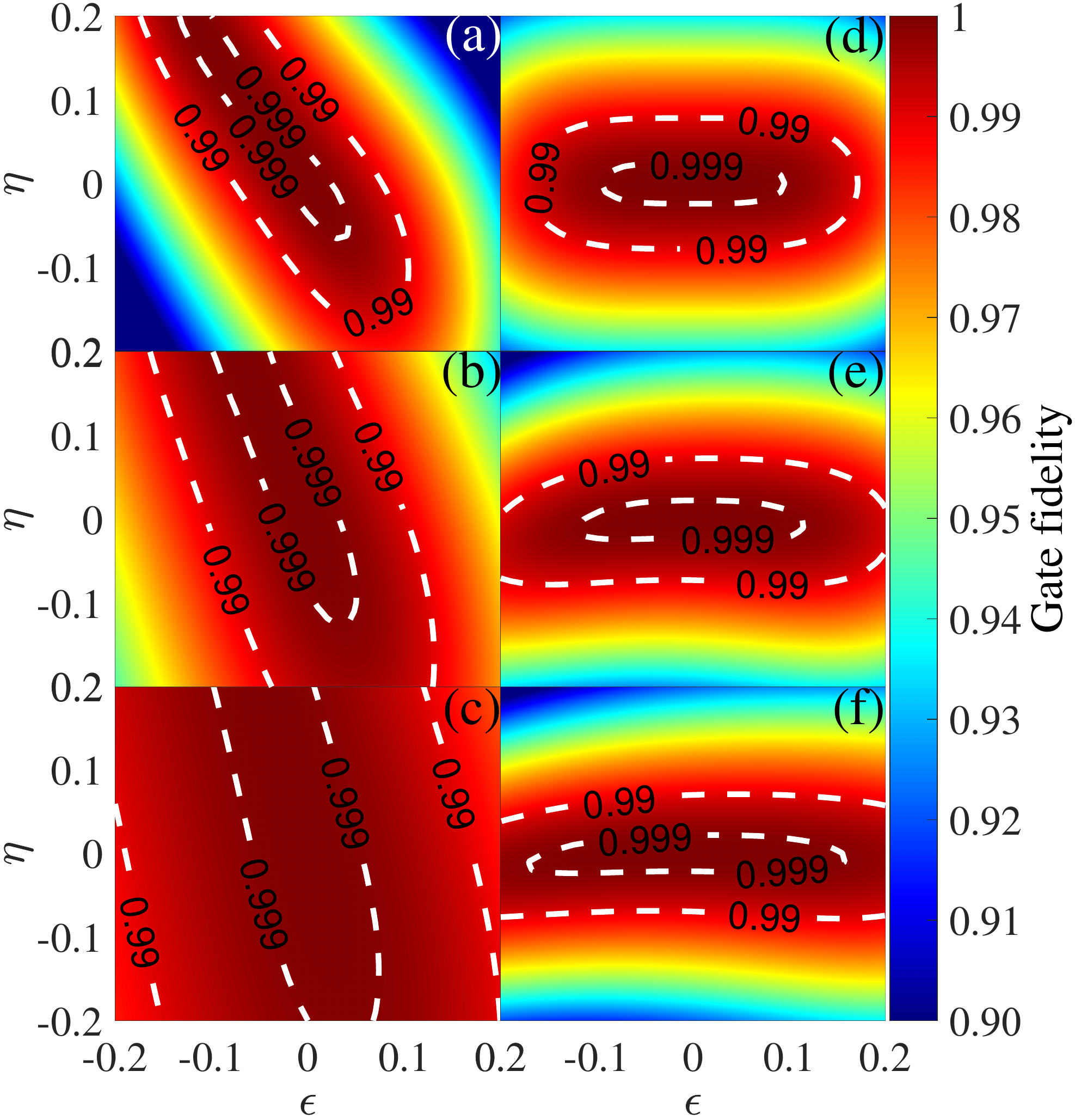}
	\caption{Comparison of the gate robustness for our scheme and the previous scheme. For the rotation gate of the $\pi$ angle (the $H$ gate), our scheme (a) has a similarly robust performance as the previous scheme (d). However, for the rotation gate of the $\pi/2$ angle (the $S$ gate), our scheme (b) has a robustness advantage over the previous scheme (e). Furthermore, for the rotation gate of the $\pi/4$ angle (the $T$ gate), our scheme (c) has a more obvious advantage in robustness compared with the previous schemes (f).}
	\label{Fig.3}
\end{figure}

Furthermore, to evaluate the gate performance under decoherence, we numerically simulate the gate fidelity as a function of the decoherence rate for the $H$, $S$, and $T$ gates, via the master equation 
\begin{equation}
    d\rho(t)/dt = i[\rho(t),\mathcal{H}(t)]+\frac{\kappa}{2}[\mathcal{L}(\sigma_1)+\mathcal{L}(\sigma_2)],
    \label{ME}
\end{equation}
where $\rho(t)$ is the density operator of the system, $\mathcal{L}(A) \!=\! 2A\rho A^{\dag}\!-\!A^{\dag}A\rho\!-\!\rho A^{\dag}A$ is the Lindbladian operator, and the decay and dephasing of the system are respectively considered as $\sigma_1\!=\!|0\rangle\langle2|\!+\!|1\rangle\langle2|$ and $\sigma_2\!=\!2|2\rangle\langle2|\!-\!|1\rangle\langle1|\!-\!|0\rangle\langle0|$. The gate fidelity is defined as $F \!=\! (1/6)\sum_{k=1}^6\langle\Psi(0)|_k U^{\dag}(\tau)\rho(\tau) U(\tau)|\Psi(0)\rangle_k$, where $|\Psi(0)\rangle_k \!=\! \{|0\rangle,|1\rangle,(|0\rangle\!+\!|1\rangle)/\sqrt{2},(|0\rangle\!-\!|1\rangle)/\sqrt{2},(|0\rangle\!+\!i|1\rangle)/\sqrt{2},(|0\rangle\!-\!i|1\rangle)/\sqrt{2}\}$ respectively. As shown in Fig. \ref{Fig.2}(b), for the previous SR-NHQC, all gates have the same decoherence performance due to constant gate time, but for our scheme, the smaller the rotation angle, the shorter the gate time, and thus the better decoherence performance.

In addition, we numerically simulate the $H$, $S$, and $T$ gates' robustness by the master equation Eq. (\ref{ME}), where the Hamiltonian considers the coupling strength error and the detuning error, i.e., $\mathcal{H}^{\epsilon,\eta}(t) = \mathcal{H}(t)+\left[\epsilon\Omega/2|2\rangle\langle b|+\text{H.c.}\right]+\eta\Omega/2|2\rangle\langle 2|$ with $\epsilon$ and $\eta\in[-0.2,0.2]$. And we set the decoherence rate $\kappa = 0$ to compare only the effect of errors on gate fidelity. As shown in Fig. \ref{Fig.3}, for the rotation gate of the $\pi$ angle (the $H$ gate), our scheme has a similarly robust performance as the previous scheme, but on the small angle rotation gates (the $S$ and $T$ gates), our scheme is significantly more robust. Due to the unavoidable decoherence effect in experiments, the gate performance of our scheme will be better, because our scheme also has an advantage in decoherence.

\begin{figure}[tbp]
	\includegraphics[width=0.9\linewidth]{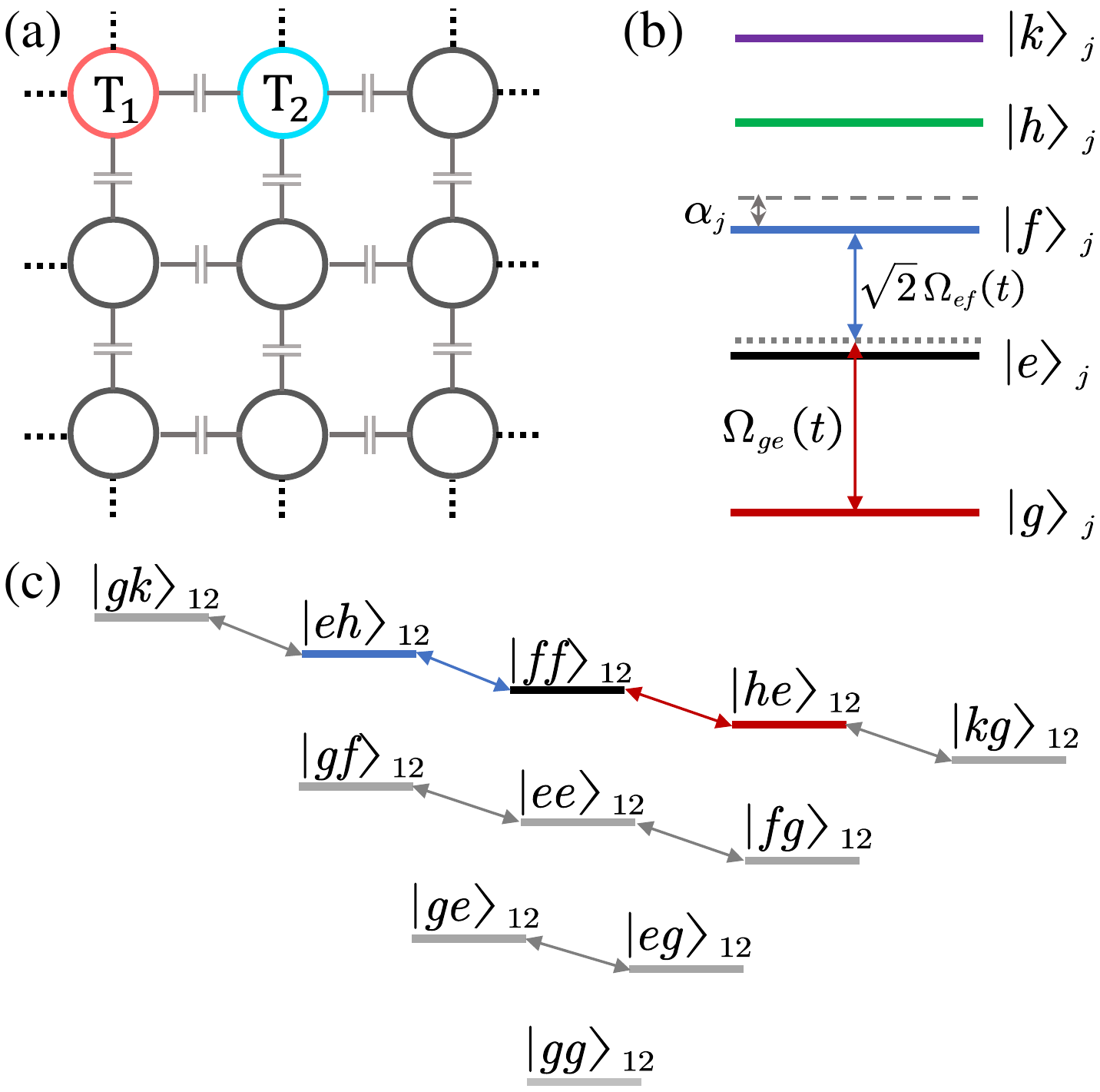}
	\caption{Illustration of the proposed implementation. (a) A superconducting lattice model composed of Transmon qubits, where each qubit ($\text{T}_j$) has a $\Xi$-type energy level structure with the anharmonic $\alpha_j$ and qubits are capacitively coupled. (b) To construct arbitrary single-qubit gates, we only add two pulses $\Omega_1(t)$ and $\sqrt{2}\Omega_2(t)$ of different frequencies to drive $|g\rangle_j\leftrightarrow|e\rangle_j$ and $|e\rangle_j\leftrightarrow|f\rangle_j$ respectively. (c) The energy level diagram of two-qubit coupling.}
	\label{Fig.4}
\end{figure}

\section{Physical Realization}
Since our scheme only needs a detuned $\Lambda$-type three-level system, it can be realized on many platforms \cite{yan2019, zhu2019, xu2018, ai2020, ai2022, li2017, zu2014, ar2014, se2017, zhou2017, is2018, na2018}. As an instance, here we introduce how our scheme can be implemented on a superconducting quantum circuit, which  is a rapidly  developing platform for the physical implementation of quantum computation. We consider a superconducting lattice model composed of Transmon qubits \cite{koch2007}, where each qubit ($\text{T}_j$) has a $\Xi$-type energy level structure with the anharmonic $\alpha_j$ and qubits are capacitively coupled, as shown in Fig. \ref{Fig.4}(a). To construct arbitrary single-qubit gates, we only add two pulses $\Omega_{ge}(t)$ and $\sqrt{2}\Omega_{ef}(t)$ with different frequencies to drive $|g\rangle_j\leftrightarrow|e\rangle_j$ and $|e\rangle_j\leftrightarrow|f\rangle_j$ respectively, as shown in Fig. \ref{Fig.4}(b), where $\Omega_{ge}(t) = \Omega_1(t)\cos{\left[\upsilon_1 (t)+\phi_1(t)\right]}$ and $\Omega_{ef}(t) = \Omega_2(t)\cos{\left[\upsilon_2 (t)+\phi_2(t)\right]}$. It is sufficient to consider four energy levels for single-qubit gates and thus the initial Hamiltonian is
\begin{align}
    \mathcal{H}_{1}(t) &\!=\! \omega_1|e\rangle\langle e|+ (2\omega_1\!-\!\alpha_1)|f\rangle\langle f|+ 3(\omega_1-\alpha_1)|h\rangle\langle h| \notag \\
    &+[\Omega_{ge}(t)+\Omega_{ef}(t)](|g\rangle\langle e|+\sqrt{2}|e\rangle\langle f|+\sqrt{3}|f\rangle\langle h| \notag \\
    &+\text{H.c.}).
\end{align}
We choose two drive frequencies as $\upsilon_1(t) \!=\! \omega_1\!+\!\Delta_1(t)$ and $\upsilon_2(t) \!=\! \omega_1\!-\!\alpha_1\!-\!\Delta_1(t)$, where $\Delta_1(t)$ is the detuning.

\begin{figure}[tbp]
	\includegraphics[width= \linewidth]{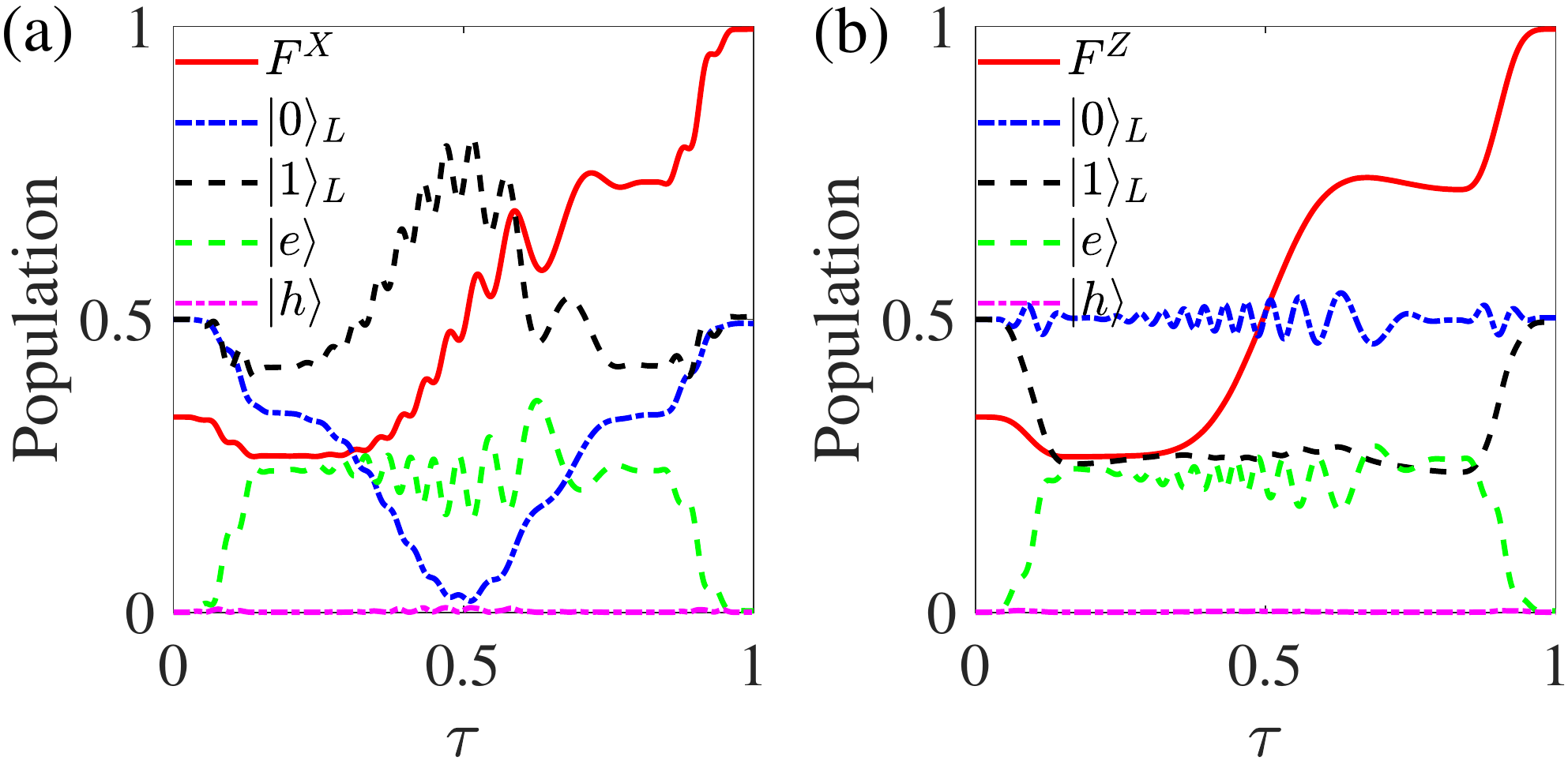}
	\caption{The state population for the $X$ gate (a) and $Z$ gate (b) with the decoherence effect, where the initial state of both gates is $(|0\rangle_L+i|1\rangle_L)/\sqrt{2}$, and the gate fidelities are $99.53\%$ and $99.52\%$ respectively.}
	\label{Fig.5}
\end{figure}

By making a unitary transformation with $\exp\{-i[(\omega_1 t \!+\! \int_0^{t'}{\Delta_1(t') dt'})|e\rangle\langle e| \!+\! (2\omega_1 \!-\! \alpha_1)t|f\rangle\langle f| \!+\! (3\omega_1-3\alpha_1)t|h\rangle\langle h|]\}$ and omitting anti-rotation wave terms, the Hamiltonian is
\begin{align}
    \mathcal{H}^I_{1}(t) &= -\Delta_1(t)|e\rangle\langle e| \notag \\
    &+\frac{1}{2}\Omega_1(t) e^{i\phi_1(t)}\left[|g\rangle\langle e|+\sqrt{2}e^{i[\alpha_1 t+2\Delta_1(t)]}|e\rangle\langle f|\right. \notag \\&\left.+\sqrt{3}e^{i[2\alpha_1 t+\Delta_1(t)]}|f\rangle\langle h|+\text{H.c.}\right] \notag \\
    &+\frac{1}{2}\Omega_2(t) e^{i\phi_2(t)}\left[e^{-i[\alpha_1 t+2\Delta_1(t)]}|g\rangle\langle e|+\sqrt{2}|e\rangle\langle f|\right. \notag \\
    &\left.+\sqrt{3}e^{i[\alpha_1 t-\Delta_1(t)]}|f\rangle\langle h|+\text{H.c.}\right],
\end{align}
where there are four leakage terms related to the anharmonic $\alpha_1$. However, due to the increasing anharmonic of the present and the derivative removal by adiabatic gate (DRAG) method \cite{motzoi2009,gambetta2011}, the four leakage terms can be well suppressed, and thus the effective Hamiltonian can be reduced to
\begin{align}
    \mathcal{H}^{eff}_{1}(t) &= -\Delta_1(t)|e\rangle_j\langle e|+\frac{1}{2}\left[\Omega_1(t) e^{i\phi_1(t)}|g\rangle_j\langle e|\right. \notag \\
    &\left.+\sqrt{2}\Omega_2(t) e^{i\phi_2(t)}|e\rangle_j\langle f|+\text{H.c.}\right].
    \label{H_1e}
\end{align}
When we choose the $\{|g\rangle,|f\rangle\}$ subspace as the single qubit logical space, i.e., $\{|g\rangle,|f\rangle\} \!=\! \{|0\rangle_L,|1\rangle_L\}$, by setting $\Omega_c(t) \!=\! \sqrt{\Omega_1^2(t)\!+\!2\Omega_2^2(t)}$, $\tan(\theta/2) \!=\! \Omega_1(t)/[\sqrt{2}\Omega_2(t)]$ and $\varphi \!=\! \phi_2(t)\!-\!\phi_1(t)\!+\!\pi$, the Eq. (\ref{H_1e}) has the same mathematical form as the Eq. (\ref{H0}), and thus the arbitrary super-robust holonomic single-qubit gates can be realized by the piecewise Hamiltonian of the Eq. (\ref{3Hc}). Through numerical simulations, we show the state populations and the gate fidelities with decoherence effect for the $X$ and $Z$ gates in Fig. \ref{Fig.5}, where the gate fidelity is $99.53\%$ for the $X$ gate, in which the decoherence effect and the leakage terms lead to the infidelity of $0.40\%$ and $0.07\%$ respectively, and the gate fidelity is $99.52\%$ for the $Z$ gate, in which the decoherence effect and the leakage terms lead to the infidelity of $0.39\%$ and $0.09\%$ respectively (see Appendix B for details).

Under the $\{|g\rangle,|f\rangle\}$ coded logical subspace, it is difficult to implement a non-trivial two-qubit gate with only two physical qubits because of the severe leakage terms. Here, we introduce a method to realize the two-qubit phase gate in the available experimental conditions with only two physical qubits and obtain the gate fidelity exceeding $99\%$. Let us use two adjacent Transmon qubits $\text{T}_1$ and $\text{T}_2$ as an example, and then the logical space for two qubits is $\{|gg\rangle_{12},|gf\rangle_{12},|fg\rangle_{12},|ff\rangle_{12}\} = \{|00\rangle_L,|01\rangle_L,|10\rangle_L,|11\rangle_L\}$. By adding a two-tone driving to $\text{T}_1$, i.e., $\omega_1(t) = \omega_1+\mu_1(t)+\mu_2(t)$ with $\mu_j(t) = \varepsilon_j\cos{\left[\upsilon'_j t+\phi'_j(t)\right]}$, the quantum system consisting of two-qubit coupling is controlled by
\begin{align}
    \mathcal{H}_2(t) &\!=\! \sum_{j=1}^2\left[\omega_j|e\rangle_j\langle e|\!+\! (2\omega_j\!-\!\alpha_j)|f\rangle_j\langle f|\right. \notag \\
    &\left.+3(\omega_j\!-\!3\alpha_j)|h\rangle_j\langle h|+(4\omega_j\!-\!6\alpha_j)|k\rangle_j\langle k|\right]  \\
    &+\mathrm{g}_{12}\!\left[(|g\rangle_1\langle e|\!+\!\sqrt{2}|e\rangle_1\langle f|\!+\!\sqrt{3}|f\rangle_1\langle h|\!+\!2|h\rangle_1\langle k|)\right. \notag \\
    &\left.\otimes(|e\rangle_2\langle g|\!+\!\sqrt{2}|f\rangle_2\langle e|)\!+\!\sqrt{3}|h\rangle_2\langle f|\!+\!2|k\rangle_2\langle h|)\!+\!\text{H.c.}\right],\notag
\end{align}
where five energy levels for each qubit are considered and the energy level diagram of a two-qubit coupling is shown in Fig. \ref{Fig.4}(c). In the interaction picture, the Hamiltonian containing the logical subspace is
\begin{align}
    \mathcal{H}_2^I(t)   =&\sqrt{2}\mathrm{g}_{12} e^{-i\Delta_{12} t} \prod_{j=1,2}{e^{-iF_j(t)}} \notag \\  
    &\times \left\{ (|01\rangle_L\langle ee|e^{-i \alpha_2 t} 
  + |ee\rangle_L\langle 10|e^{i \alpha_1t} ) \right. \notag \\
    &  +\sqrt{3}[|eh\rangle_L\langle 11|e^{i(\alpha_1-2\alpha_2)t} \notag \\
    &    \left.+|11\rangle_L\langle he|e^{i(2\alpha_1-\alpha_2)t}] \right\} +\text{H.c.},
    \label{HI_2}
\end{align}
where $\Delta_{12} = \omega_1-\omega_2$ and $F_j(t) = \beta_j \sin{\left[\upsilon'_j t+\phi'_j(t)\right]}$ with $\beta_j = \varepsilon_j/\upsilon'_j$. And there are two three-energy structures, i.e., $\{|01\rangle_{L},|ee\rangle_{12},|10\rangle_{L}\}$ and $\{|eh\rangle_{12},|11\rangle_{L},|he\rangle_{12}\}$, that can be used to achieve the two-qubit phase gate. When driving one of the three-level structures, the other becomes the source of the leakage terms. Due to the larger coupling coefficient for higher energy levels, we construct the two-qubit phase gate in the subspace of $\{|eh\rangle_{12},|11\rangle_{L},|he\rangle_{12}\}$ by selecting two drive frequencies as $\upsilon'_1 = \Delta_{12}-2\alpha_1+\alpha_2+\Delta_2$ and $\upsilon'_2 = \Delta_{12}-\alpha_1+2\alpha_2-\Delta_2-\upsilon'_1$, where $\Delta_2$ is the detuning.

Applying Jacobi–Anger expansion and omitting the high-frequency oscillation terms, we can obtain
\begin{align}
    \mathcal{H}_2^I(t) &=\sqrt{2}J_{11}\mathrm{g}_{12}e^{i\left[\phi'_1(t)+\phi'_2(t) \right]}|01\rangle_L\langle ee|e^{i(\alpha_2\!-\!\alpha_1\!-\!\Delta_2) t} \notag \\
    &-\sqrt{2} J_{10} \mathrm{g}_{12}e^{-i\phi'_1(t)}|10\rangle_L\langle ee|e^{i(\alpha_1\!-\!\alpha_2\!-\!\Delta_2) t} \notag \\
    &+\sqrt{6} J_{11} \mathrm{g}_{12}e^{i\left[\phi'_1(t)+\phi'_2(t) \right]}|eh\rangle_L\langle 11|e^{-i\Delta_2 t} \notag \\
    &-\sqrt{6} J_{10} \mathrm{g}_{12}e^{-i\phi'_1(t)}|he\rangle_L\langle 11|e^{-i\Delta_2 t}+\text{H.c.},
\end{align}
where $J_{mn} \!=\! J_m(\beta_1) J_n(\beta_2)$ and there are two low-frequency leakage terms dominated by $\left|\alpha_1-\alpha_2\right|$, which leads to serious leakage from the logical space, and thus it is hard to realize  high-fidelity holonomic two-qubit gate with only two physical qubits. However, within current experimental achievable $\mathrm{g}_{12}$ and $\left|\alpha_1-\alpha_2\right|$, we can adjust parameters $\beta_j$ to minimize the influence of the leakage terms. Under the unitary transformation with $\exp\{i\Delta_2 t(|ee\rangle_{12}\langle ee|+|ff\rangle_{12}\langle ff|)\}$, the effective Hamiltonian after omitting leakage terms is
\begin{align}
    \mathcal{H}_2^{eff}(t) &= \Delta_2|11\rangle_L\langle 11|+\left[\mathrm{g}_1 e^{-i\phi'_1(t)}e^{-i\varphi'}|eh\rangle_L\langle 11|\right. \notag \\
    &\left.-\mathrm{g}_2 e^{-i\phi'_1(t)}|he\rangle_L\langle 11|+\text{H.c.}\right],
    \label{H_2e}
\end{align}
where $\varphi' = -2\phi'_1(t)-\phi'_2(t)$, $\mathrm{g}_1 = \sqrt{6}J_1(\beta_1) J_1(\beta_2)\mathrm{g}_{12}$ and $\mathrm{g}_2 = \sqrt{6}J_1(\beta_1) J_0(\beta_2)\mathrm{g}_{12}$. By setting $\mathrm{g} = \sqrt{\mathrm{g}_1^2+\mathrm{g}_2^2}$ and $\tan(\theta'/2) = \mathrm{g}_1/\mathrm{g}_2$, the Eq. (\ref{H_2e}) has the same mathematical form as the Eq. (\ref{H0}). With the piecewise Hamiltonian of the Eq. (\ref{3Hc}), we can add an arbitrary phase in the logical state $|11\rangle_L$, i.e.
\begin{equation}
    U'(\tau) = e^{-i\gamma'}|11\rangle_L\langle11|,
\end{equation}
where $\gamma' = \pi[1-\cos{\chi(\tau_1)}]$. When we set parameters as $\{\theta',\varphi',\gamma'\} = \{0,0,\pi\}$, in the logical space of $\{|00\rangle_L,|01\rangle_L,|10\rangle_L,|11\rangle_L\}$, the $CZ$ gate can be obtained as
\begin{equation}
    U'(\tau) = \mathrm{diag}\ (1,1,1,-1).
\end{equation}

\begin{figure}[tbp]
	\includegraphics[width=0.9\linewidth]{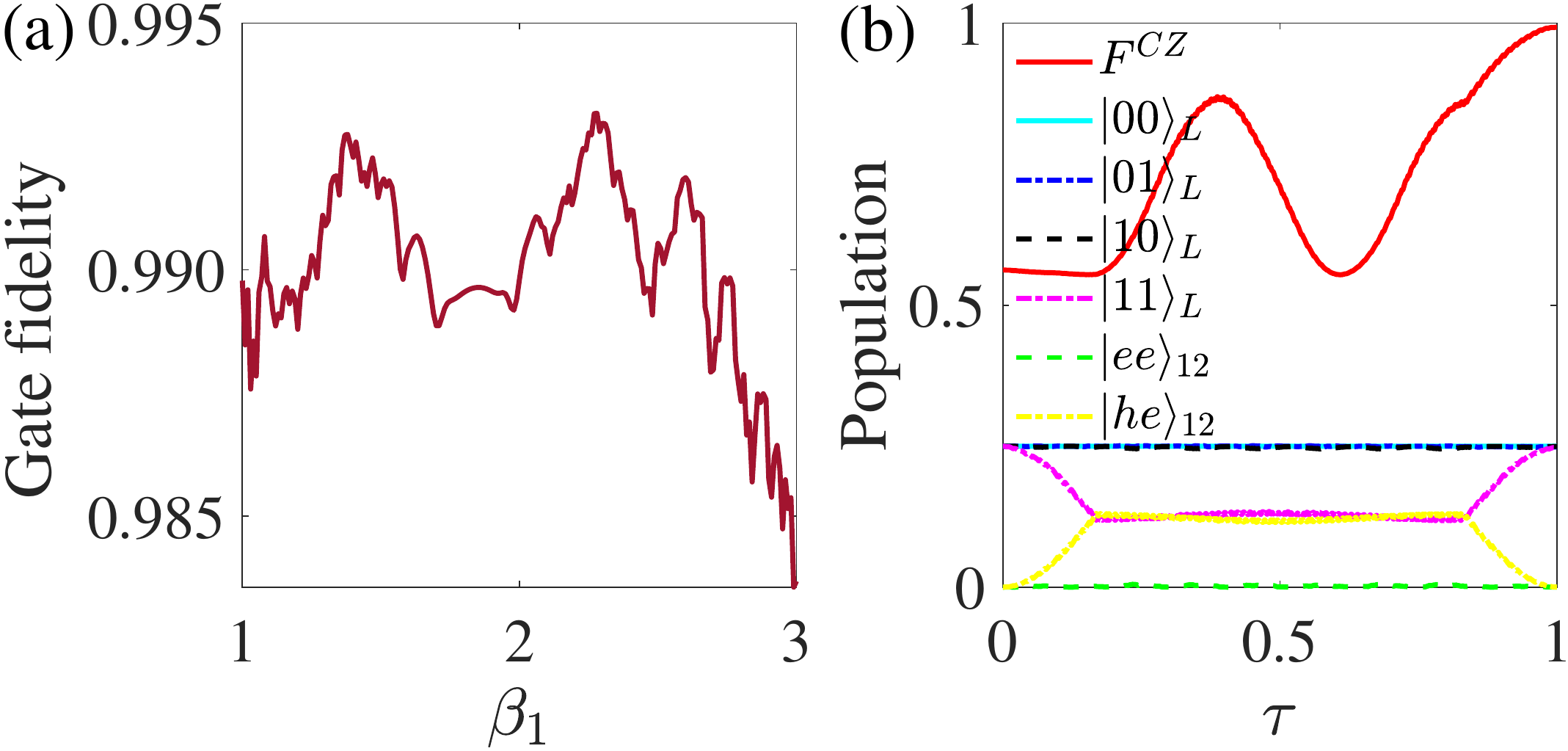}
	\caption{(a) The gate fidelity as a function of the parameter $\beta_1$ to search for the optimal operation point minimizing the influence of the leakage terms. (b) The state population and the gate fidelity with decoherence effect for the $CZ$ gate, where the initial state is set as $(|00\rangle_L+|01\rangle_L+|10\rangle_L+|11\rangle_L)/2$ and the gate fidelity is $99.30\%$.}
	\label{Fig.6}
\end{figure}

In order to investigate the feasibility of our proposed two-qubit gate, we numerically simulate the state population and the gate fidelity with the decoherence effect for the $CZ$ gate via the master equation Eq. (\ref{MEs}), where the Hamiltonian Eq. (\ref{HI_2}) is used. Meanwhile, the decay operator and the dephasing operator are changed to $S_1=\sum_{j=1,2}{(|g\rangle_j\langle e|+\sqrt{2}|e\rangle_j\langle f|+\sqrt{3}|f\rangle_j\langle h|+2|h\rangle_j\langle k|)}$ and $S_2=\sum_{j=1,2}{(|e\rangle_j\langle e|+2|f\rangle_j\langle f|+3|h\rangle_j\langle h|+4|k\rangle_j\langle k|)}$ respectively. The experimental parameters are set as $\mathrm{g}_{12}=2\pi\times 5\ \mathrm{MHz}$, $\beta_2 \!=\! 0$ , $\alpha_1 = 2\pi\times 300\ \mathrm{MHz}$, $\alpha_2 = 2\pi\times 380\ \mathrm{MHz}$, $\Delta_{12} = 2\pi\times 10^3\ \mathrm{MHz}$, and $\kappa=2\pi\times 2\ \mathrm{KHz}$, which can be obtained experimentally. We numerically simulate the gate fidelity as a function of the parameter $\beta_1$ to search for the optimal operation point minimizing the influence of the leakage terms, as shown in Fig. \ref{Fig.6}(a). And the two-qubit gate fidelity is defined as $F \!=\! (1/16)\sum_{k=1}^{16}\langle\Psi(0)|_k U^{\dag}(\tau)\rho(\tau) U(\tau)|\Psi(0)\rangle_k$, where $|\Psi(0)\rangle_k \!=\! \{|g\rangle_1,(|g\rangle_1\!-\!i|f\rangle_1)/\sqrt{2},(|g\rangle_1\!+\!|f\rangle_1)/\sqrt{2},|f\rangle_1\} \otimes \{|g\rangle_2,(|g\rangle_2\!-\!i|f\rangle_2)/\sqrt{2},(|g\rangle_2\!+\!|f\rangle_2)/\sqrt{2},|f\rangle_2\}$ respectively. Subsequently, setting the optimal parameter $\beta_1 \!=\! 2.3$, we show the state population for the $CZ$ gate with the initial state $(|00\rangle_L+|01\rangle_L+|10\rangle_L+|11\rangle_L)/2$, and the gate fidelity is $99.30\%$, as shown in Fig. \ref{Fig.6}(b). So far, in the superconducting lattice model composed of Transmon qubits, our scheme can construct arbitrary single-qubit gates and a non-trivial two-qubit gate, i.e., forming a set of universal quantum gates, and thus our scheme provides another option for realizing quantum computation in superconducting platforms.

\section{Conclusion}
In summary, we propose an accelerated SR-NHQC scheme based on a $\Lambda$-type three-level system with detuning. Compared with the previous SR-NHQC, our scheme only needs the three-segment Hamiltonian to construct arbitrary rotation gates, which requires the six-segment Hamiltonian in the previous scheme. Besides, our gate time depends on the rotation angle, and the smaller the rotation angle, the shorter the gate time, and thus our scheme has a great improvement in decoherence performance due to the shortening of our gate time. Moreover, our scheme is more robust than the previous scheme for the coupling strength error and the detuning error, especially in small-angle rotation gates. In addition, we give a comprehensive physical realization method on a 2D superconducting quantum circuit.  From our numerical simulations, the single-qubit gate fidelity can be above $99.50\%$ and the two-qubit  $CZ$ gate fidelity is $99.30\%$.  Therefore, our scheme provides a fast and implementable alternation for the SR-NHQC.

\begin{acknowledgements}
This work was supported by the Key-Area Research and Development Program of GuangDong Province (Grant No. 2018B030326001), the National Natural Science Foundation of China (Grant No. 12275090) and Guangdong Provincial Key Laboratory (Grant No. 2020B1212060066).
\end{acknowledgements}

\appendix
\section{The Hamiltonian of the previous SR-NHQC}
The previous scheme \cite{li2021} needs a resonant $\Lambda$-type three-level system controlled by
\begin{equation}
    \mathcal{H}_L(t) = \Omega(t)|2\rangle\langle b|+\text{H.c.}
\end{equation}
In order to satisfy the super-robust condition, the Hamiltonian is divided into the following six segments
\begin{align}
    \mathcal{H}_L^1(t) &=  \frac{1}{2}i\Omega(t) e^{i\zeta_0}|2\rangle\langle b|+\text{H.c.}, \quad t\in[0,\tau_1], \notag \\
    \mathcal{H}_L^2(t) &=  \frac{1}{2}i\Omega(t) e^{i(\zeta_0+\pi/2)}|2\rangle\langle b|+\text{H.c.}, \quad t\in(\tau_1,\tau_2], \notag \\
    \mathcal{H}_L^3(t) &=  \frac{1}{2}i\Omega(t) e^{i\zeta_0}|2\rangle\langle b|+\text{H.c.}, \quad t\in(\tau_2,\tau_3], \notag \\
    \mathcal{H}_L^4(t) &=  -\frac{1}{2}i\Omega(t) e^{i(\zeta_0+\gamma)}|2\rangle\langle b|+\text{H.c.}, \quad t\in(\tau_3,\tau_4], \notag \\
    \mathcal{H}_L^5(t) &=  -\frac{1}{2}i\Omega(t) e^{i(\zeta_0-\pi/2+\gamma)}|2\rangle\langle b|+\text{H.c.}, \quad t\in(\tau_4,\tau_5], \notag \\
    \mathcal{H}_L^6(t) &=  -\frac{1}{2}i\Omega(t) e^{i(\zeta_0+\gamma)}|2\rangle\langle b|+\text{H.c.}, \quad t\in(\tau_5,\tau],
\end{align}
and requires
\begin{align}
    &\int_0^{\tau_1}{\Omega(t)}dt = \pi/2, \quad t\in[0,\tau_1], \notag \\
    &\int_{\tau_1}^{\tau_2}{\Omega(t)}dt = \pi, \quad t\in(\tau_1,\tau_2], \notag \\
    &\int_{\tau_2}^{\tau_3}{\Omega(t)}dt = \pi/2, \quad t\in(\tau_2,\tau_3], \notag \\
    &\int_{\tau_3}^{\tau_4}{\Omega(t)}dt = \pi/2, \quad t\in[\tau_3,\tau_4], \notag \\
    &\int_{\tau_4}^{\tau_5}{\Omega(t)}dt = \pi, \quad t\in(\tau_4,\tau_5], \notag \\
    &\int_{\tau_5}^{\tau}{\Omega(t)}dt = \pi/2, \quad t\in(\tau_5,\tau].
\end{align}
For a fair comparison with our scheme, we set $\Omega(t) = \Omega$, and thus the total gate time $\tau = 4\pi/\Omega$.

\begin{figure}[tbp] 
	\includegraphics[width= \linewidth]{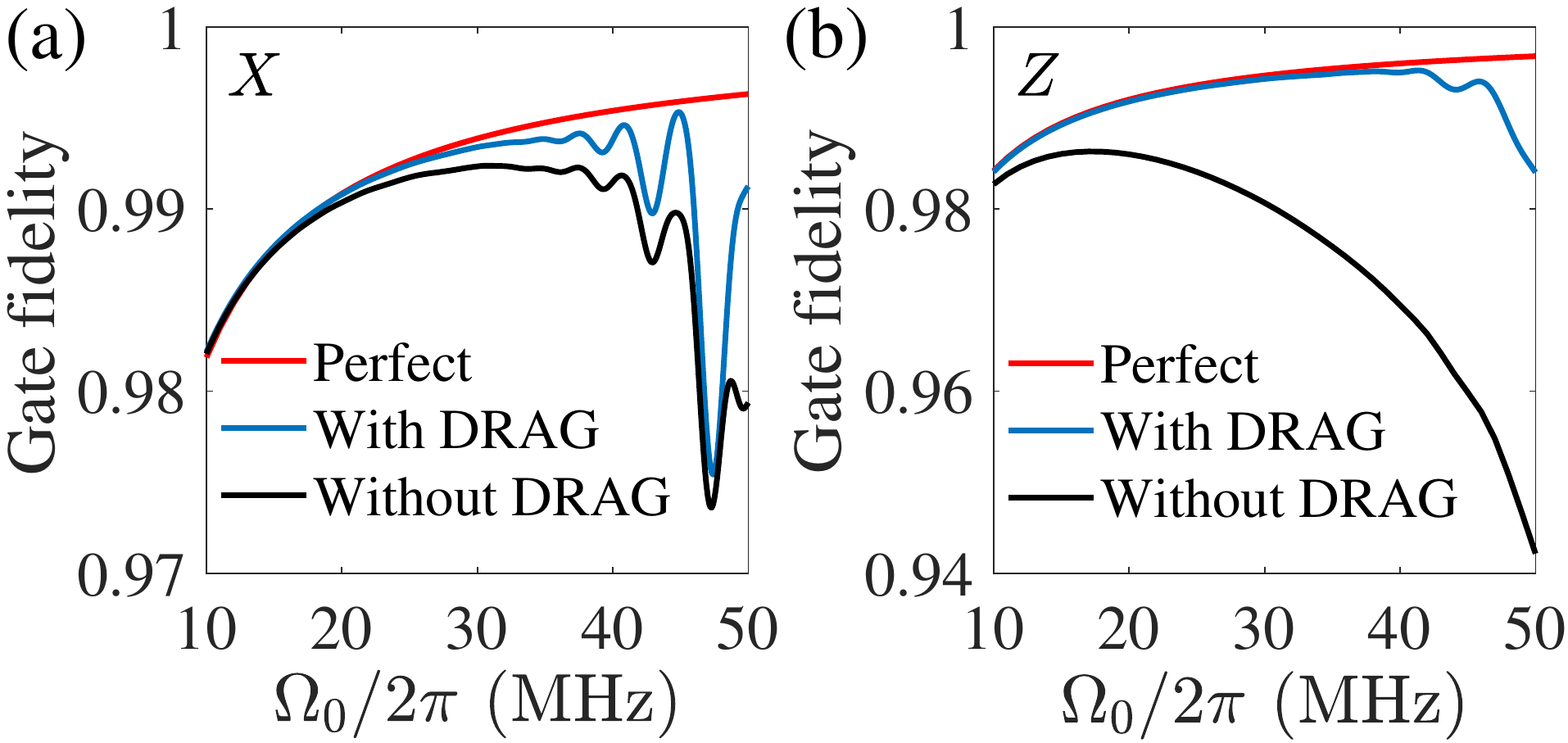}
	\caption{The gate fidelity as a function of $\Omega_0$ for $X$ gate (a) and $Z$ gate (b) with the decoherence rate $\kappa=2\pi\times 2\ \mathrm{KHz}$, where the red line represents the result of the ideal Hamiltonian $\mathcal{H}_1^{eff}(t)$, the black line represents the result of the non-ideal Hamiltonian $\mathcal{H}_1^I(t)$, and the blue line represents the result of the non-ideal Hamiltonian $\mathcal{H}_1^I(t)$ with DRAG.}
	\label{Fig.7}
\end{figure}

\section{The single-qubit gates with DRAG in superconducting circuits}

Due to the low anharmonic of the Transmon qubits, the DRAG correction \cite{motzoi2009,gambetta2011} is required for gate operation in superconducting circuits, i.e. $\tilde{\Omega}_1(t) \!=\! \Omega_1(t)\!-\!ip_1\dot{\Omega}_1(t)/\alpha_1$, $\tilde{\Omega}_2(t) \!=\! \Omega_2(t)\!-\!ip_2\dot{\Omega}_2(t)/\alpha_1$ and $\tilde{\Delta}_1(t) \!=\! \Delta_1(t)\!+\!p_3\Omega^2_1(t)/(2\alpha_1)\!+\!p_4\Omega^2_2(t)/(2\alpha_1)$, where $\{p_j\}$ are the parameters used for optimization and $\Omega_j(0) \!=\!\Omega_j(\tau) \!=\! 0$ is necessary for the DRAG. By setting the drive amplitudes $\Omega_1(t) \!=\! \Omega_0 \sin{(\theta/2)}\sin^2{(\pi t/\tau)}$ and $\Omega_2(t) \!=\! \Omega_0 \cos{(\theta/2)}\sin^2{(\pi t/\tau)}/\sqrt{2}$, we numerically
simulate the gate fidelity as a function of $\Omega_0$ for the $X$
 and $Z$ gates, via the master equation
\begin{equation}
    d\rho(t)/dt = i[\rho(t),\mathcal{H}(t)]+\frac{\kappa}{2}[\mathcal{L}(S_1)+\mathcal{L}(S_2)],
    \label{MEs}
\end{equation}
where $\rho(t)$ is the density operator of the system, $\mathcal{L}(A) \!=\! 2A\rho A^{\dag} \!-\! A^{\dag}A\rho \!-\! \rho A^{\dag}A$ is the Lindbladian operator, and the decay and dephasing of the system are respectively considered as $S_1 \!=\! |g\rangle\langle e|\!+\!\sqrt{2}|e\rangle\langle f|\!+\!\sqrt{3}|f\rangle\langle h|$ and $S_2 \!=\! |e\rangle\langle e|\!+\!2|f\rangle\langle f|\!+\!3|h\rangle\langle h|$. According to the current experimental conditions, we set the anharmonic $\alpha_1 \!=\! 2\pi\times 300\ \mathrm{MHz}$ and the decoherence rate $\kappa \!=\! 2\pi\times 2\ \mathrm{KHz}$. In order to find the DRAG's best operation point and investigate sources of infidelity, we numerically simulate the $X$ gate and $Z$ gate by utilizing the ideal Hamiltonian $\mathcal{H}_1^{eff}(t)$, the non-ideal Hamiltonian $\mathcal{H}_1^I(t)$, and the non-ideal Hamiltonian $\mathcal{H}_1^I(t)$ with DRAG respectively. Meanwhile, for the $X$ gate, the DRAG's optimization parameters $\{p_j\}$ are set as: $p_1\!=\!1.8$, $p_2\!=\!-1$, $p_3\!=\!-0.4$, $p_4\!=\!-0.3$, and for the $Z$ gate, the optimization parameters are set as: $p_1\!=\!0$, $p_2\!=\!1.6$, $p_3\!=\!0$, $p_4\!=\!-2.8$. Moreover, as shown in Fig. \ref{Fig.7}, the best operation points for the $X$ and $Z$ gates are $\Omega_0 \!=\! 2\pi\times 45\ \mathrm{MHz}$ and $\Omega_0 \!=\! 2\pi\times 41\ \mathrm{MHz}$ respectively. Subsequently, we investigate the state populations and the gate fidelities with decoherence effect for the $X$ and $Z$ gates by applying the parameters of the best operation point. As shown in Fig. \ref{Fig.5}, for the $X$ gate, the gate fidelity is $99.53\%$, in which the decoherence effect and the leakage terms lead to the infidelity of $0.40\%$ and $0.07\%$ respectively, and for the $Z$ gate, the gate fidelity is $99.52\%$, in which the decoherence effect and the leakage terms lead to the infidelity of $0.39\%$ and $0.09\%$ respectively.


\end{document}